\documentclass[useAMS,usenatbib]{mn2e}

\input epsf

\usepackage{graphicx}
\usepackage{txfonts}
\usepackage{epsfig}
\usepackage{times}
\usepackage[margin=0.5in, top=1in, bottom=1in, a4paper, pdftex, dvips]{geometry}

\title[Determining weighted power spectra]{A new efficient method for determining weighted power spectra: Detection of low frequency solar p-modes by analysis of BiSON data}
\author[S. T. Fletcher et. al.]
{S. T. Fletcher$^1$, A-M. Broomhall$^2$\thanks{E-mail:amb@bison.ph.bham.ac.uk}, W. J. Chaplin$^2$, Y. Elsworth$^2$ and R. New$^1$\\
$^1$ Materials Engineering Research Institute, Faculty of Arts,
Computing, Engineering and Science, Sheffield Hallam University,\\
Howard Street, Sheffield, S1 1WB, UK  \\
$^2$School of Physics and Astronomy, University of Birmingham,
Edgbaston, Birmingham, B15 2TT, UK \\}
\begin{document}

\maketitle

\begin{abstract}
We present a new and highly efficient algorithm for computing a power spectrum made from evenly spaced data which combines the noise-reducing advantages of the weighted fit with the computational advantages of the Fast Fourier Transform (FFT). We apply this method to a 10-year data set of the solar p-mode oscillations obtained by the Birmingham Solar Oscillations Network (BiSON) and thereby uncover three new low-frequency modes. These are the $\ell$=2, $n$=5 and $n$=7 modes and the $\ell$=3, $n=$7 mode. In the case of the $\ell$=2, $n$=5 modes, this is believed to be the first such identification of this mode in the literature. The statistical weights needed for the method are derived from a combination of the real data and a sophisticated simulation of the instrument performance. Variations in the weights are due mainly to the differences in the noise characteristics of the various BiSON instruments, the change in those characteristics over time and the changing line-of-sight velocity between the stations and the Sun. It should be noted that a weighted data set will have a more time-dependent signal than an unweighted set and that, consequently, its frequency spectrum will be more susceptible to aliasing.
\end{abstract}


\section{Introduction}\label{sec_intro}

Low-angular-degree (low-$\ell$) solar acoustic (p) modes travel deep into the solar interior and as such can be used as probes of the solar core. Results from various completed and ongoing observational programmes have revealed a very rich spectrum of low-$\ell$ p-modes in a frequency range of around 1.0 to 5.5 mHz. An important and ongoing goal in helioseismology has been the detection of acoustic modes with ever lower frequencies. This is because low frequency acoustic modes are extremely long lived and so once they are detected their frequencies can be determined to very high precision. This enhances their diagnostic potential as probes of the solar interior. Unfortunately the effects of increasing solar granulation background and decreasing amplitude of the modes as one moves lower in frequency makes detection very difficult.

The Birmingham Solar Oscillations Network (BiSON) has been collecting low-$\ell$ data for over 3 decades and as such provides an excellent opportunity to detect very low frequency acoustic modes. In order to obtain as continuous observations as possible BiSON comprises six stations, separated in longitude around the Earth, each of which houses a resonance scattering spectrometer (RSS). However, as the network was not built all at once, but rather expanded over time from just a single station to the current six, the RSS's were not produced to the same specifications. As such, the noise level of the data is different at each station. This means that the combined time-series generated from joining together data from all the stations has a noise function that varies with time.

Matters are further complicated by the fact that any Doppler shift measurements made by an RSS will have uncertainties and noise characteristics that depend on the line-of-sight velocity between the instrument and the Sun \citep[see][]{Chaplin2005a}. The noise level changes through the day as the Earth spins and through the year as the Earth orbits the Sun.

This time dependence of the noise suggests that, if we wish to make best use of the BiSON data and produce power spectra with the lowest possible noise, we should weight the data. If the noise level in the power spectra can be significantly reduced by such a method it may be possible to increase the signal-to-noise ratio (S/N) by a sufficient amount to detect more low frequency modes. The S/N needs to be assessed for each particular case because it also depends on the aliasing that results from multiplying the signal by the weight. Over the course of this paper we shall describe how a weighted power spectrum for the BiSON data was produced using a new computationally efficient method. This is important when looking at very long time series as the traditional sine wave fitting approach would take a very long time to compute (of the order of a few weeks even when using a high performance computing cluster). We go on to present results on how much the S/N is improved and discuss the detection of new low frequency modes.

\section{The Data}\label{sec_data}

In addition to using real BiSON data we have also used simulations to inform us about any improvements in S/N and the likelihood of detecting new modes. However, for the simulated data to be useful we need the noise characteristics to follow closely the real data. The most effective way of doing this is to simulate the solar absorption line and then run through the acquisition and analysis of the data in the same way as is done with the real solar observations. In order to explain how this is done we start by briefly explaining how the real data are collected.

At all six stations a resonance scattering spectrometer (RSS) \citep[see][]{Brookes1978} is used to measure the Doppler velocity shift of the 770-nm D1 potassium absorption line. In each case incident solar light is passed through a cell containing a vapour of potassium atoms and photons of the appropriate energy are resonantly scattered from the incident beam. Ideally, detectors placed at right angles to the incident beam record only scattered photons. The absorption cross section of atoms in the cell is much narrower than that of the solar lines because the temperature in the cell is much lower than that of the Sun and there is no rotational broadening. This means only light from a relatively narrow band of the solar absorption line is detected.

The vapour cell is placed in a magnetic field, which causes Zeeman splitting to occur. The single line is thus split into a multiplet with separations dependent on the field strength. The splitting alters the atomic states in such a way that atoms will interact with circularly polarized light. The blue-shifted transition is sensitive to one hand of circular polarization whilst the red-shifted transition is sensitive to the other. The splitting also improves the sensitivity of the RSS by moving the passbands out onto the wings of the Fraunhofer line where the slope is greatest and hence any given line shift will result in a greater change in measured intensity.

A linear polariser and an electro-optic quarter-wave plate are used to circularly polarise the incident light and, by switching quickly between one-handedness and the other, it is possible to measure the light intensity in the blue wing, $I_b$ and red wing, $I_r$, almost simultaneously. From these measurements a ratio, $R$, is formed which gives a near-linear proxy for the velocity shift of the solar line:
\begin{equation}
R = \frac{I_b - I_r}{I_b + I_r}. \label{Rshort}
\end{equation}
To obtain velocity measurements of the solar oscillations, $R$ is plotted against the line-of sight station velocity (with respect to the Sun) and a third-order polynomial function is fitted. The oscillations signal is then recovered by subtracting $R$ from the polynomial and is calibrated using the fitted gradient of $R$ versus the station velocity \citep[see][]{Elsworth1995}.

This is an idealised situation and in reality background offsets and instrumental noise will affect the intensity measurements used to form $R$. A more realistic expression is thus given by \citep{Chaplin2005a}:
\begin{equation}
R = \frac{I_b^{res}+I_b^{non}+I_b^{elec}+i_b-I_r^{res}-I_r^{non}-I_r^{elec}-i_r} {I_b^{res}+I_b^{non}+I_b^{elec}+i_b+I_r^{res}+I_r^{non}+I_r^{elec}+i_r}
\label{Rlong}
\end{equation}
where $I^{res}$ is the desired contribution from resonantly scattered light and $I^{non}$, $I^{elec}$ and $i$ are background sources due to non-resonantly scattered light, electronic offsets and noise respectively.

The measurements of the solar oscillations are collected at each station each day and combined to form a continuous time series with a duty cycle of around 86\% (for 2007 and 2008). However, there are often occasions when two or even three stations are collecting data at the same time. In these cases the usual approach is to use the data with the lowest noise characteristics when producing the time series, although it is also possible to take an average of the data sets.

In order to properly characterise how the various noise sources propagate through this complex system we use a realistic simulation of the BiSON data designed to include many of the instrumental noise effects in addition to the actual solar oscillations. The first step in the simulation process is to estimate $I^{res}$ by fitting a Gaussian profile to an observed line obtained using Doppler velocity observations of the centre of the solar disc. The observations were taken by the Themis solar telescope located at Iza$\tilde{\textrm{n}}$a, Tenerife [private communication with Rosaria Simoniello, 2008]. The effects of solar rotation, limb darkening, Doppler imaging and image rotation are all taken into account in order to generate a simulated line similar to that seen by the RSS's at each of the BiSON stations \citep[see][]{Broomhall2009a}. The use of this more realistic line shape is an enhancement on earlier simulation work reported by \cite{Chaplin2005a}.

The `operating point' on the simulated line is evaluated from the Doppler shift due to the changing line of sight velocity of the Sun to each station (approximate values for the gravitational red shift and convective blue shift are included). Artificial velocity oscillations can then be added. Intensity `measurements' are made for $I_b^{res}$ and $I_r^{res}$ in the same way as with real data. At this point estimates of the various noise sources can easily be included. Finally, the ratio can be formed and analysed in the same way as with real data.

In order to test the benefits of a weighted power spectrum we used a 10-year (3650-day) stretch of BiSON data covering the period January 1st 1997 to December 31st 2007. To correspond with this we also created a set of simulated data of the same length. In order to better observe any reductions in the background we chose not to add any additional noise due to a granulation-like background into the simulated data. However, the instrumental noise was based on that of real BiSON data, in order to make the weighting functions similar. The simulated data contained an oscillations signal created in the manner described in \cite{Jimenez2008} which has been generated specifically to mimic the real solar signal.

\section{Efficient Weighted Power Spectrum}\label{sec_efficient}

The traditional method of determining a power spectrum from a time series of data with known weights is to perform a sine-wave fitting (SWF) analysis \citep[see e.g.,][]{Kjeldsen1992,Frandsen1995,New2009}. If all the measurements have equal weight, then the power spectrum computed for the evenly-spaced BiSON data takes the same form, whether one uses the computationally efficient Fast Fourier Transform (FFT), the SWF or the definition adopted by Scargle for the periodogram \citep{Scargle1982}. In previous publications we have always assumed equal weighting of BiSON data and always calculated the power spectrum using the FFT.

The SWF is far more computationally expensive than simply taking the FFT since the fits must be calculated for each separate frequency. In fact, to analyse a 10-year time series of BiSON data using an SWF method would take many weeks even when using a high performance computing cluster. As such, this makes the repeated calculations needed for scientific analysis very difficult. Therefore, in studying the effect of applying weighting to BiSON data, we have first developed a new method whereby we substitute certain summations made during the SWF process with far more computationally efficient FFT's.

Here we describe the main differences between our new method and the traditional SWF method. We leave a detailed derivation for Appendix 2. For a standard SWF approach, one fits:
\begin{equation}
A_i \sin(2\pi f_i t_k) + B_i \cos(2\pi f_i t_k),
\end{equation}
to the data, $y_k$, at each frequency, $f_i$, where $t_k$ is the time of the $k$th observation. $A_i$ and $B_i$ can be evaluated by least squares fitting. The appropriate expressions are \citep[Appendix 1;][]{Kjeldsen1992,Frandsen1995}:
\begin{equation}
A_i = \frac {\sum w_k y_k c_{ik} \sum w_k s_{ik} c_{ik} - \sum w_k y_k s_{ik} \sum w_k c_{ik}^2} {(\sum w_k s_{ik} c_{ik})^2 - \sum w_k c_{ik}^2 \sum w_k s_{ik}^2},
\end{equation}
and
\begin{equation}
B_i = \frac {\sum w_k y_k s_{ik} \sum w_k s_{ik} c_{ik} - \sum w_k y_k c_{ik} \sum w_k s_{ik}^2} {(\sum w_k s_{ik} c_{ik})^2 - \sum w_k c_{ik}^2 \sum w_k s_{ik}^2},
\end{equation}
where $s_{ik}=\sin(2\pi f_i t_k)$, $c_{ik}=\cos(2\pi f_i t_k)$ and $w_k$ is the statistical weight of the $k$th measurement. The power spectrum can then be obtained by evaluating $A_i^2 + B_i^2$  as a function of $i$. Computing the power spectrum in this manner takes considerably longer than computing it via an FFT because each unique summation given in the expression for $A_i$ and $B_i$ must be evaluated at each frequency. However, as we show in Appendix 2, by using certain properties of the Fourier Transform and by making certain (very good) approximations it is possible to rewrite the expressions for $A_i$ and $B_i$ in terms of an FFT and a summation over the weights.
\begin{equation}
A_i = \frac{-N \textrm{Im}[\textrm{FFT}(w_k y_k)]}{\sum(w_k/2)}
\end{equation}
and
\begin{equation}
B_i = \frac{N \textrm{Re}[\textrm{FFT}(w_k y_k)]}{\sum(w_k/2)}.
\end{equation}
We have checked both the SWF and FFT methods using short time series and showed that both returned almost identical results (i.e., to within 10$^{-6}$), for a variety of different realistic weighting functions. It should be noted that this method can only be applied to power spectra that have been made from evenly spaced data.

\section{Determining the Weighting}\label{sec_method}

In \cite{New2009} it was shown that by using the known noise characteristics of the Global Oscillations at Low Frequencies (GOLF) instruments on board the Solar and Heliospheric Observer (SOHO) spacecraft, it was possible to weight the time-series and hence produce a weighted power spectrum. This resulted in an improved signal-to-noise ratio (defined as the ratio of the heights of the modes to the background and hereafter denoted S/N). The same principle can be applied to BiSON data although determining the noise associated with the points in the time series is a little more difficult. This is because the noise characteristics change both on a daily and yearly basis due to a change in the line-of-sight velocity, and because the time series is made up from observations from six different stations each of which have different noise characteristics and which can also change over time due to mechanical wear, breakdown and parts being replaced.

There are two possible methods of tracking the underlying noise characteristics of the various BiSON instruments. The first is to make some estimate based on measurements taken directly from the real data, and the second is to simulate the observation and analysis of the data as accurately as possible and predict the noise based on those simulations. In this work we combine both of these methods in order to derive as accurate a representation of the noise characteristics as possible.

For the first approach we attempt to estimate the noise characteristics of the time series (and hence determine a weighting) by looking at the daily power spectrum for each station. The envelope of power associated with the strongest acoustic oscillations extends from about 1.3 mHz to 5.5 mHz. Due to the exponential drop off of the solar granulation background the power at frequencies above the regime of acoustic modes will be almost entirely due to instrumental noise. While for frequencies below this regime the mean power will be a combination of the instrumental noise and the solar granulation background. We expect only the instrumental noise to vary significantly over time but we still choose to determine the noise characteristics from the low frequency regimes. This is because it is known that the instrumental noise increases at lower frequencies and it is unknown whether this increase will always be proportional to the high frequency instrumental noise. Additionally the main aim of reducing the noise level and increasing the S/N is to detect weak low frequency acoustic modes, so weighting the time series according to the low frequency noise would seem the most logical choice.

\begin{figure}
\centerline{\includegraphics[width=3.5in]{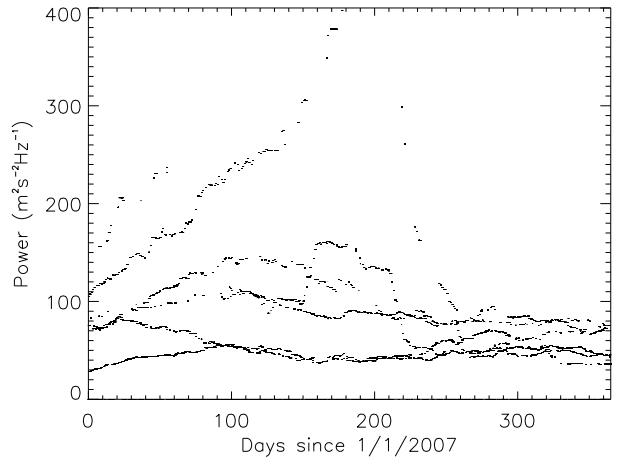}}\caption{Variation in the amplitude of the noise for a BiSON time series for the year 2007. The fact that 6 independent traces can be made out highlights the fact that the noise level of the various stations remain relatively stable on a day-to-day basis, different stations tend to have very different noise levels (by up to about a factor of 8) and that data from all the stations are used in order to make up the full time series.} \label{TimeSeriesNoiseGran}
\end{figure}

Once the daily mean noise values have been determined they can be matched to each point in the time series by tracking which data point comes from which station for each day. The main drawback of this approach is one of precision. Each day's data only lasts for around 12-13 hours at best (significantly less during winter months and/or if there is poor weather). Since we average only over a limited range of the power spectrum the variation in estimated noise from one day to the next for any particular station can be quite large. This can be alleviated somewhat by smoothing the daily values for each station over a number of days (say 60). However this approach runs the risk of giving disproportionably higher weightings to bad days of data that occur during a stretch of good days and vice versa. A second but related problem is that this method cannot be used to track how the noise varies throughout the day. Fig.~\ref{TimeSeriesNoiseGran} shows the noise in the 0.8-1.3mHz range associated with each data point for a BiSON time series for the year 2007.

The second approach to predicting the noise variations is to use simulated data created in the manner described in section~\ref{sec_data}. By adding in different noise sources into the simulator it is relatively easy to show how the change in the operating point affects the level of instrumental noise on a daily and yearly basis \citep[see][]{Chaplin2005a,Fletcher2009c}. However it is much more difficult to predict the magnitude and type of noise for each RSS at the different stations. Therefore in creating the simulated data we scale the noise to approximately match that of the best performance seen in the real data for each of the stations. We also use the amount by which the instrumental noise varies throughout the years in order to predict the most likely type of noise.

The big advantages of artificial data are that we can simulate the noise in isolation to the modes and also add in an oscillations signal with known characteristics. The simulator can also be used to show how the noise varies throughout the day. Therefore, we choose to combine the two methods described above by superimposing the predicted daily variations onto the day-by-day noise measurements from the real data and hence determine a noise value for each individual data point. In summary we can use the measurements of the station noise levels to set out long term weight variations and simulations of daily effects to set out diurnal weight variations. However, as we go on to explain in section~\ref{sec_results}, including the daily effect seem to have only a negligible impact on the resulting S/N of the weighted power spectrum.

\section{Analysis}\label{sec_analysis}

In this section we discuss the various ways in which we have analysed the power spectrum. For our initial attempt at weighting the time series we have assumed the weights, $w_k$, to be the inverse of the noise, where the noise is determined for each data point using the method outlined in the previous section. The weight is set to zero where there are gaps in the time series.

The best way of determining how much of an improvement is seen in the S/N when weighting the data, is to actually compare the S/N of the individual modes in power spectra with and without weighting, and the results of doing this are given in the following section. However, it is also a useful exercise to try to make a prior prediction of the improvement in the S/N.

The prediction is based on considerations of the signal and noise levels that are expected in the unweighted and weighted power spectra. In the unweighted case the time series is constructed from measurements from the various BiSON instruments. While all such instruments are expected to measure the same signal level the noise level for each instrument is different (strictly, each measurement is different, since the noise level depends also on the line of sight velocity). According to Parseval's Theorem the sum of power in a power spectrum is proportional to the variance of the time series used to compute the power spectrum, and the constant of proportionality is unity for a common definition of the Discrete Fourier Transform\footnotemark[1].

\footnotetext [1]{ Strictly, the sum of the power in the power spectrum is related to the sum of the power in the time series by:
\begin{equation}\sum_{i=0}^{N-1} {\vert X(i)\vert^2}={N \alpha^2} \sum_{k=0}^{N-1} {x(k)^2},
\end {equation}
where $\alpha$ is defined by the expression for the DFT. The general form of the DFT is:
\begin {equation} \textrm{DFT}x(t)=X(i)=\alpha \sum_{k=0}^{N-1} {x(k) e^{-2\pi j n k/N}}.\end{equation}
Common choices for $\alpha$ are unity \citep[see, for example, page 97 of][]{Brigham1988} or $\frac{1}{N}$ \citep[see, for example, page 608 of][]{Press2007}. The latter is used by the \textsc{IDL} language, and we have adopted this definition throughout the paper. Making this choice for $\alpha$ means that the multiplier on the right hand side of Equation 8 is $\frac{1}{N}$, hence the statement in the main text relating the sum of the power in the power spectrum to the variance of the time series.}

Hence, using this definition, the expected total noise power, $b^u$,  in an unweighted spectrum obtained from a time series of $N$ points is:
\begin{equation}
b^u = \frac{\sum{b_k}}{N}. \label{NoisePowerUnw}
\end{equation}
where, $b_k$ is the variance attributable to a given point.

When we weight the data we are essentially transforming a time series which is the original multiplied by the weights (see the argument at the end of Appendix 2). In that case the standard deviation of each point in the weighted time series is $w_k \sqrt{b_k}$, and the total noise power, $b^w$ obeys:
\begin{equation}
b^w = \frac{\sum{(b_k w_k^2)}}{N}, \label{NoisePower}
\end{equation}

In this work we define the weights as the inverse of the noise relative to the lowest noise level of any measurement, $b_1$:
\begin{equation}
w_k = \frac{b_1}{b_k}. \label{WeightDefine}
\end{equation}
Hence, we can write:
\begin{equation}
b^w = b_1\frac{\sum{w_k}}{N} \label{NoisePowera}
\end{equation}
and, using Eqns.~\ref{NoisePowerUnw} and ~\ref{WeightDefine},
\begin{equation}
b^u = b_1\frac{\sum{\frac{1}{w_k}}}{N}. \label{NoisePowerb}
\end{equation}

We now go on to consider the signal. For a continuous sinusoidal signal of amplitude $a$, the total signal strength in the power spectrum will be $a^2/2$ (as required by Parseval's theorem). Hence, using Eqn.~\ref{NoisePowerb} we see that the S/N for the unweighted case will be given by:
\begin{equation}
S/N^u=\frac{a^2N}{2b_1\sum{\frac{1}{w_k}}}.\label{S/NUnw}
\end{equation}

In the weighted case, each point is multiplied by the appropriate weight and, assuming that variations in weight occur slowly compared to the periods of p modes of interest (so that the weight is constant over the period of the mode), the total signal power is reduced by a factor of $\sum{w_k^2}/N$. Hence, the overall power, $S$, in the power spectrum would be:
\begin{equation}
S = \frac{a^2}{2} \left(\frac{\sum{w_k^2}}{N}\right). \label{SignalPower}
\end{equation}

However, not all of that power will be contained within the main peaks of the spectrum as some will be removed into, for example, diurnal sidebands. We can see the effect that weighting the time series has on the peaks in the power spectrum by making use of the convolution theorem. As the time-series we work with in the weighted case is the product of the weighting function $w$ with the data $y$ (see the argument at the end of Appendix 2), the transform will be the convolution of the transform of $w$ with that of $y$. Firstly, we note that the transform of the purely sinusoidal $y$ will just result in peaks of height $a/2$ at frequencies of $\pm 1/P$, where $P$ is the period of the sinusoid. Secondly, while $w_k$ (and hence its transform) is likely to be complicated, it takes the form of the sum of a complicated function, $F$, centred on zero with a d.c. level, $D$, which is given by:
\begin{equation}
D=\frac{1}{N}\sum{w_k}.
\end{equation}
Hence, making use of the linearity of Fourier Transforms, the transform of $w$ will be the sum of transform of $D$, which is just a peak of height $\frac{1}{N}\sum{w_k}$, with that of $F$.
So the complete transform of $w y$ will contain peaks of height $\frac{a}{2N}\sum{w_k}$, at frequencies of $\pm 1/P$ and the total power in peaks at frequencies of $\pm 1/P$ in the power spectrum will be $H$, given by:
\begin{equation}
H = \frac{a^2}{2} \left(\frac{\sum{w_k}}{N}\right)^2. \label{SignalHeight}
\end{equation}

The S/N in the weighted case, $S/N^w$, is determined by dividing Eqn.~\ref{SignalHeight} by Eqn.~\ref{NoisePowera}, thus:
\begin{equation}
S/N^w=\frac{a^2\sum{w_k}}{2Nb_1}.\label{S/NW}
\end{equation}

In order to obtain separate values of the S/N in both the unweighted and weighted cases we would need to know $a^2$. However if we take the ratio of the two S/N's the $a^2$ terms cancel and we can obtain an estimate of the fractional improvement in the S/N for the weighted spectra compared with the unweighted one without the need to know $a^2$, i.e. we can divide Eqn.~\ref{S/NW} by Eqn.~\ref{S/NUnw} to obtain:
\begin{equation}
\frac{S/N^w}{S/N^u}=\frac{\sum{\frac{1}{w_k}}\sum{w_j}}{N^2}.\label{S/Nratio}
\end{equation}

Firstly we note that this is always greater than 1 (see Appendix 3 for a proof). I.e. with the assumptions made, weighting is always advantageous. Secondly, for the real BiSON case we find that we should see a improvement in the S/N of around 50\% if we weight the data. However, this is only a rough estimate since the amplitudes of the modes are not completely constant over time and the fact that the values of $b_k$ can only be estimated (see Section~\ref{sec_method}). It turns out that the expected improvement based on these equations actually gives an overestimate of the true improvement we see (see section~\ref{sec_results}). Even so, Eqs.~\ref{NoisePowerUnw} through \ref{S/Nratio} do help us better understand the effect of weighting the data.

\begin{figure}
\centerline{\includegraphics[width=3.0in]{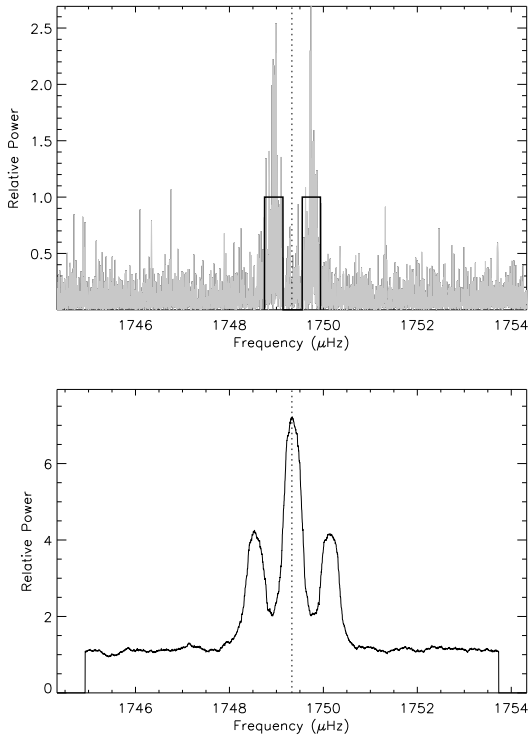}}\caption{Top panel: Small slice of the power spectrum of a 3650-day BiSON time series centered on the $\ell$=1, $n$=11 mode. The spectrum has been divided by the underlying background. Overlaid on the spectrum is the smoothing function designed to smooth the spectra over the two multiplet components. Bottom panel: The resulting smoothed spectrum.} \label{ModesConvol}
\end{figure}

A more accurate method for estimating the improvement in the S/N is to look at the power spectrum directly. It is possible to fit the background at low and high frequencies in order to get an estimate of the power in the noise without the need to use Eqs.~\ref{NoisePower}~or~\ref{NoisePowera} and thus removing the need to estimate $b_k$. This can be done for both the weighted and unweighted cases and can then be combined with Eq.~\ref{SignalHeight} to obtain another estimate of how much the S/N increases when using weighted data.

Although these predictions are useful in order to help discern the benefits from weighting the data, it is clear that we must look at the individual S/N of the modes to get a full picture. One way of doing this is to fit each of the modes in the spectrum using a standard peak bagging approach. However, the modes at very low frequencies have extremely long lifetimes and consequently are very narrow in the frequency domain and are often difficult to fit. We therefore chose to employ a strategy whereby we smoothed the data (see below) and simply read off the maximum peak height. However, we also take into account that the modes are rotationally split. Hence, the smoothing function for the $\ell$ = 0 modes is a simple box car with a width equal to twice the expected linewidth of the mode. Whereas for the $\ell$ = 1 modes the smoothing function is a double box car with a spacing between them (measured from the centre of the boxes) equal to the rotational splitting. The $\ell$ = 2 and 3 modes have 3 and 4 box cars respectively to coincide with all the rotationally split multiplets. An example of this multiple box-car type smoothing, for an $\ell$ = 1 mode, can be seen in Fig.~\ref{ModesConvol}. The big advantage of this approach is that, as well as smoothing the data, we also combine all the power from all the multiplets into a large peak at the central frequency of the mode. This has a distinct advantage when looking for modes with a weak S/N. This method is similar to the $m$-averaging method developed by \cite{Salabert2008}.

It is also important to think about how to scale the data in order to compare more directly the power spectrum determined from the weighted and unweighted time series. Given that the aim of weighting the data is to reduce the background and hence improve the S/N, then the best way of scaling is to make the power of the signal (i.e the modes) the same in both cases. This can be done by fitting and removing the background in the two spectra and then summing the power over the region of the modes (say 1.3-5.5 mHz), and dividing the one total by the other in order to obtain a scaling factor. Or, alternatively, we can fulfil Parsevals's theorem by dividing both the unweighted and weighted spectra by $\sum{w_k^2}/N$ (where in the case of the unweighted data, $w_k$ would only take values of one or zero), although, as stated previously, this method will only be accurate if the amplitude of the modes do not vary much with time (i.e., if the signal is a pure sinusoid).

\section{Results}\label{sec_results}

We now go on to present the effects of weighting the BiSON data on the power spectra and give the results on how much the S/N of the modes can be improved. In section~\ref{subsec_results_sim} we look at the simulated data where we have a very clear understanding of how the noise function varies throughout the day and year. The simulated data also have the advantage that the input frequencies for the modes are known so we can calculate the S/N where the modes should be occurring even if those modes are too weak to be seen above the level of the noise. In section~\ref{subsec_results_real}, we give the results of weighting the real data.

\subsection{Simulated Data}\label{subsec_results_sim}

\begin{figure*}
\centerline{\includegraphics[width=5.0in]{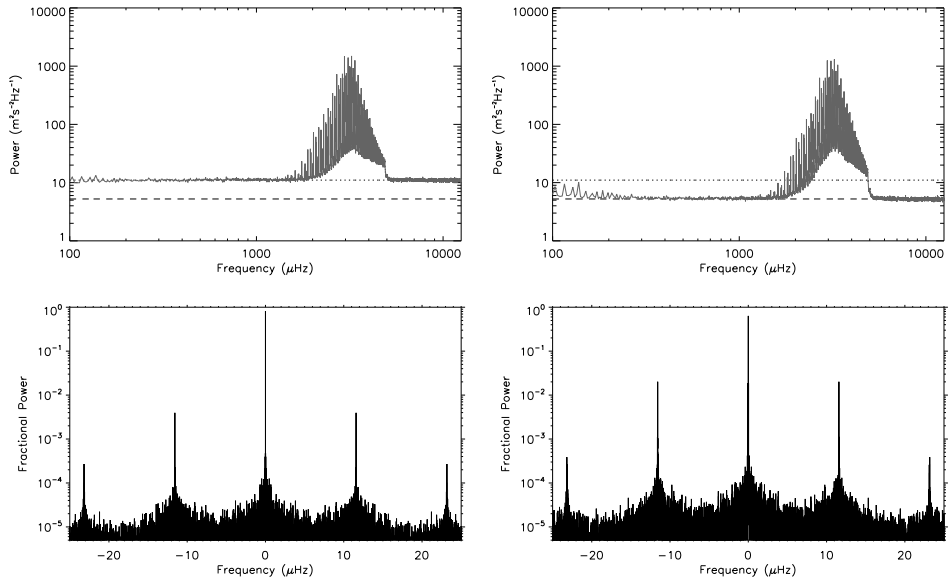}}\caption{Smoothed power spectra of the unweighted (top left) and weighted (top right) simulated time series. The data has been scaled so that the power in the modes is the same in both plots, however, more of that power will be aliased out into side bands in the weighted case. Dotted and dashed lines shows the underlying background in the unweighted and weighted cases respectively. Bottom panels show the spectral windows for the two power spectra. } \label{PowSpecSimON}
\end{figure*}

The top two panels in Fig.~\ref{PowSpecSimON} show the power spectrum of the unweighted and weighted time series of the simulated data. We have scaled the data by dividing by the mean of the squared weights in order to make the power in the modes the same in both plots (see Eq.~\ref{SignalPower}). We could also have chosen to scale the plots to give the same peak heights using Eq.~\ref{SignalHeight} or to give equal backgrounds using Eq.~\ref{NoisePowera}. The plot shows that there is an overall reduction in the background level of the weighted spectrum. We remind the reader that for the simulated data we chose not to include a granulation like signal in order to more clearly observe any reduction in the background. (Hence the scales of  Fig.~\ref{TimeSeriesNoiseGran} and Fig.~\ref{PowSpecSimON} are not directly comparable.) The bottom two panels give the corresponding spectral windows and show that the weighted spectrum has significantly more power removed from the main peak than the unweighted data. This means that the improvement due to the decrease in the background will be somewhat offset by the smaller height of the central peak.

\begin{figure}
\centerline{\includegraphics[width=3.0in]{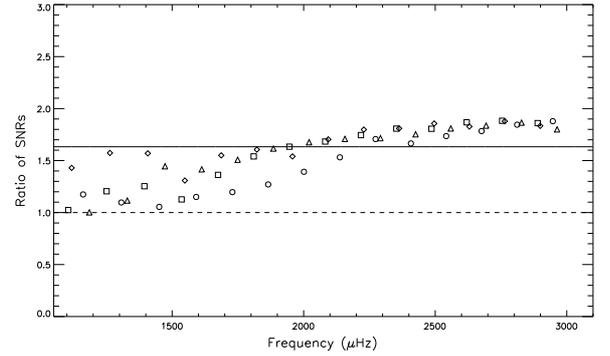}}\caption{Ratio of the S/N at known mode frequencies for the simulated weighted and unweighted data. The symbols represent different $\ell$, with $\ell$=0 given by diamonds, $\ell$=1 by triangles, $\ell$=2 by squares and $\ell$=3 by circles. The solid line shows the `predicted' improvement in S/N (see text) and a dotted line is given at unity to aid the eye.} \label{S/NRSim} \end{figure}

Using the mode-like smoothing function discussed in Section~\ref{sec_analysis}, we are able to compare the S/N of the individual modes. Note that since we know exactly where the modes should occur for the simulated spectrum, we take the S/N of the highest peak within a very restricted range of the known frequency ($\pm0.05\mu$Hz). Fig.~\ref{S/NRSim} shows the S/N of the modes in the weighted spectrum divided by the S/N of the modes in the unweighted spectrum. The different symbols indicate different $\ell$ and the solid line indicates the `predicted' improvement in the S/N for the simulated data calculated by dividing the reduction in the power of the main mode peak by the reduction in the background.

The plot shows that we can expect an improvement in the S/N of up to around 60\% and the measurement of the S/N of the individual modes broadly agrees with this. However it is noticeable that at mid to high frequencies where the mode peaks are much stronger and broader the S/N improvement seems to be somewhat better than expected. This was found to be an effect of the mode width. Looking closely at the bottom panels of Fig.~\ref{PowSpecSimON} it can be seen that there is power in the bins surrounding the central peak. The effect of convolving a wide peak with such a spectral window is to give a peak in the power spectrum whose height is somewhat greater than if both functions had been `narrower'. In the case of the weighted data there is more power surrounding the main central peak than in the unweighted data. Hence the effect on the peak's height is greater and we see an increase in the ratios of the S/Ns. The effect of this is further increased by the fact that we are averaging over the power in the modes when applying the mode-like smoothing function described in section~\ref{sec_analysis}. In contrast at low frequencies the improvement in the S/Ns is less than expected. This reflects the fact that the mode heights fall significantly at low frequencies and thus some modes may still not be seen above the background level (hence we will not be able to measure the true improvement in the S/N). Also, since the data is being weighted by the noise level at low frequencies, it is likely that as well as reducing the background level there will also be some reduction in the power of the low frequency modes.

In Section~\ref{sec_method} we explained how the noise function that the weighting is based on could be determined. This noise function varies both throughout the day and throughout the year due to the changing line-of-sight velocity between the stations and the Sun and also due to the different noise characteristics of the various instruments. Tracking the differences between instruments and other long term variations can be done by measuring the noise in the power spectra from daily time series from each station. However daily variations can not be measured in this way. Instead we must use the simulator to predict the variations and superimpose these onto the mean values for each day. However, the difference between using a weighting function with daily variations compared to one without turned out to be very small. Fig.~\ref{S/NRSim} gives the S/N of individual modes when fixed values are used throughout the day. Using a varying value improved these results by less than 1\%. This suggests that when weighting the real data it is not important to take into account the impact of the daily variations of the noise. This is because the absolute magnitude of the daily variation will be proportional to the mean noise over the day. Therefore, the daily variation will only be of a similar order to the difference in noise between the stations when the noise itself is large. However, when the noise is large the weighting is small and so the overall impact of the daily variation on the power spectrum ends up being small.

In the preamble of this section we also discussed that we chose to weight the time series as the inverse of the noise. This seems to be the best option. When taking the inverse of the square of the noise or the inverse of the square root of the noise the improvement in the S/N is not as good.

In order to help confirm the detection of modes statistical tests are often employed. One such method is to calculate the probability of any particular peak in the spectrum being solely due to noise \citep[see][]{Chaplin2002}. In order to do this we need to know the underlying noise distribution. For p-mode power spectra made from long uninterrupted time series this is known to to be negative exponential (i.e., $\chi^2$ with 2 degrees of freedom). Simulations were performed and showed that such a distribution remains valid even when applying the weighting function to the time series\footnotemark[1].

\footnotetext [1]{Strictly, the use of the $\chi^2$ with 2 degrees of freedom distribution is limited to  situations in which the noise in the time series is stationary, which, in practice for BiSON data, it never is. In the unweighted case the noise changes because BiSON data are taken by different instruments. In the weighted case, both noise and signal change. Therefore, pragmatically, before carrying out peak detection methods that employ statistical tests, we always check the underlying distribution.}

Therefore, the probability, $p$, of getting, by chance, a spike of power, $\zeta$ in any particular bin is:
\begin{equation}
p(\nu) = \exp\left(-\frac{\zeta_\nu}{\langle\zeta_\nu\rangle}\right), \label{prob1}
\end{equation}
where $\langle\zeta_\nu\rangle$ is an appropriate average background level. From this the probability, $P$, of there being at least one peak greater than or equal to $\zeta$ over a range of $N$ bins can be calculated:
\begin{equation}
P = 1-(1-p(\nu))^N \label{prob2}
\end{equation}
The value of $N$ must be carefully chosen by the user as its value will affect the calculation of $P$. Discussion of the best value to use has already been given in both \cite{Chaplin2002} and \cite{Broomhall2007}. Here we choose a value of $N$ equal to the number of bins in a 100 $\mu$Hz slice of our spectra. This is the same frequency range used in \cite{Broomhall2007}.

Tests can be performed in order to check for the occurrence of power over a range of bins (since the power in the modes will generally be spread over many bins given long time series) and for power arranged in the various mode like structures (i.e the rotational splitting patterns for $\ell >$ 0). As discussed earlier we haven taken a slightly different approach and smooth the peaks using a mode-like smoothing function. Therefore, we can simply apply Eq.~\ref{prob2} for the highest peak after we have taken the convolution. However, it should be noted that in calculating $p$, the fact that we are averaging over a number of bins must be accounted for. It can be shown that calculating the probability of having a power, $s$, averaged over $N_{av}$ bins for a $\chi^2$ distribution with $D$ degrees of freedom is equivalent to finding the probability of having power, $s$, in a single bin for a $\chi^2$ distribution with $DN_{av}$ degrees of freedom \citep[see][]{Appourchaux2004}.

\begin{table}
\centering \caption{Table giving the frequencies, $\nu$, for the simulated data, of the known modes for low $n$ and the probabilities, $P$, of the power at those frequencies occurring, by chance, over a range of 100 $\mu$Hz. Peaks which are strong enough to have only a low probability of occurring by chance ($P<$ 0.1) are printed in bold. Modes which are identified by this means in the weighted spectrum only are denoted by an asterisk.}
\begin{tabular}{cccccc}
\hline $\ell$ & n & $\nu_u$ ($\mu$Hz) & $P_u$ & $P_w$ \\
\hline
 0 &  7 & 1118.02 &         1.00  &         1.00  &   \\
 0 &  8 & 1263.49 &         1.00  & \textbf{0.00} & * \\
 0 &  9 & 1407.47 &         0.42  & \textbf{0.00} & * \\
 0 & 10 & 1548.42 & \textbf{0.00} & \textbf{0.00} &   \\
 0 & 11 & 1686.74 & \textbf{0.00} & \textbf{0.00} &   \\
 1 &  7 & 1185.60 &         1.00  &         1.00  &   \\
 1 &  8 & 1329.64 & \textbf{0.00} & \textbf{0.00} &   \\
 1 &  9 & 1472.92 & \textbf{0.00} & \textbf{0.00} &   \\
 1 & 10 & 1612.60 & \textbf{0.00} & \textbf{0.00} &   \\
 1 & 11 & 1749.33 & \textbf{0.00} & \textbf{0.00} &   \\
 2 &  6 & 1105.25 &         1.00  &         1.00  &   \\
 2 &  7 & 1250.81 &         1.00  &         0.77  &   \\
 2 &  8 & 1394.61 & \textbf{0.00} & \textbf{0.00} &   \\
 2 &  9 & 1535.93 & \textbf{0.00} & \textbf{0.00} &   \\
 2 & 10 & 1674.58 & \textbf{0.00} & \textbf{0.00} &   \\
 3 &  6 & 1161.74 &         1.00  &         1.00  &   \\
 3 &  7 & 1306.85 &         0.35  & \textbf{0.09} & * \\
 3 &  8 & 1451.01 & \textbf{0.08} & \textbf{0.00} &   \\
 3 &  9 & 1591.42 & \textbf{0.00} & \textbf{0.00} &   \\
 3 & 10 & 1729.14 & \textbf{0.00} & \textbf{0.00} &   \\
\hline \label{TableSim}
\end{tabular}
\end{table}

Table~\ref{TableSim} gives $P$ at the stated frequencies (which are the known input values for the modes) for both the unweighted and weighted cases. A value of $P$ close to unity indicates there is high probability that the level of power in the tested bin is due solely to the noise distribution. In the case of simulated data, where we know a mode should be present, this indicates the S/N of this mode is too low to be seen above the noise level. When $P$ is close to zero, there is only a small chance that that level of power is due to noise. A value of $P<$ 0.1 (given in a bold typeface in the table) means there is less than 10\% chance of the peak being due solely to noise and as such usually warrants being investigated further.

For regions of the spectrum where there is known to be a mode there are three occasions where $P$ is below the 0.1 threshold in the weighted data but not in the unweighted data (these modes are denoted by an asterisk in Table~\ref{TableSim}). Conversely there are no instances where the opposite is true. This suggests that not only is there a clear increase in the S/N when using the weighted power spectrum, but also that this increase may allow us to uncover more yet unidentified modes in the real BiSON data.

It is useful to estimate the value of the S/N at which we would expect the weighted power spectrum to give a detection when the unweighted spectrum does not. We estimated this by gradually increasing the power of the $\ell$ = 0 mode at 1118.02 $\mu$Hz in the simulated data and seeing whether or not we obtained a detection in both the weighted and unweighted cases. We made 20 realizations for each mode power in order to get an average value of the S/N. From this test we ascertained that an average S/N of a little above 2.0 was required in order for a single mode to be detected. Therefore, if a mode in the weighted spectrum has a S/N of about 2.0 we expect to get a positive detection whilst in the unweighted case the S/N is likely to be only about 1.7 (see Fig.~\ref{S/NRGran}), and a detection is not expected.

\subsection{Real Data}\label{subsec_results_real}

\begin{figure*}
\centerline{\includegraphics[width=5.0in]{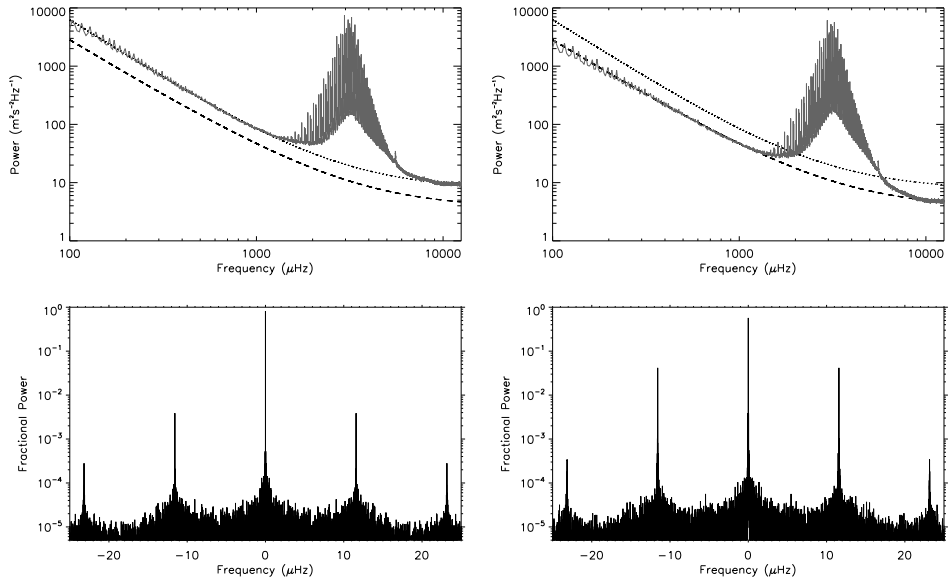}}\caption{Smoothed power spectra of the unweighted (top left) and weighted (top right) real BiSON time series. The data has been scaled so that the power in the modes is the same in both plots, however, more of that power will be aliased out into side bands in the weighted case. Dotted and dashed lines shows fits to the background in the unweighted and weighted cases respectively. Bottom panels show the spectral windows for the two power spectra.}
\label{PowSpecGran}
\end{figure*}

Fig.~\ref{PowSpecGran} shows the power spectrum and spectral windows of a real 10-year BiSON time series for both the weighted and unweighted data. Again we have scaled the plots so that the power in the modes are the same. The background signal, $n$, as a function of frequency, $\nu$, was parameterised and fitted using a power-law model \citep{Harvey1985}. A two-component power law was used, i.e.,
\begin{equation}
n(\nu) = \frac{\sigma_1^2}{1+(\nu\tau_1)^a} + \frac{\sigma_2^2}{1+(\nu\tau_2)^b} + c \label{Harvey}
\end{equation}
where $\tau_1$ and $\sigma_1$ represent the time constant and standard deviation respectively of the granulation signal, $\tau_2$ and $\sigma_2$ give the same parameters for the meso-and super-granulation component and $a$ and $b$ are the respective power law exponents, although we chose to fix $b$ with a value of 2. While we cannot be sure that a Harvey model will still be the best model for the background in the weighted case, it can still be used as a guide to the eye to show the improvement. However, we note that due to the weighting process the fit to the background is unlikely to give reliable estimates for $\tau_2$ and $\sigma_2$.

\begin{figure*}
\centerline{\includegraphics[width=5.0in]{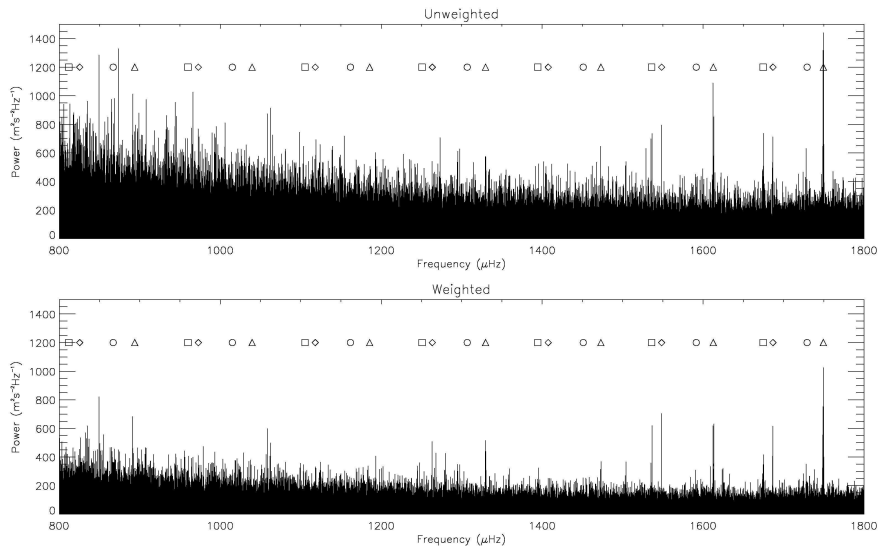}}\caption{A close up view of the power spectra of the unweighted (top) and weighted (bottom) real BiSON time series over the region 800 - 1800 $\mu$Hz . The data has been scaled in the same way as in Fig 5. but has not been smoothed. There are clearly more modes visible in the weighted spectrum compared with the unweighted.}
\label{PowSpecModesGran}
\end{figure*}

The fits to the background clearly show that there is a significant reduction in the background level in the weighted case (but of course, even if our process were perfect, could only reduce the noise level to that of the lowest obtained in the full data set). However, even though the real data were weighted according to the noise in the low frequency region of the spectrum, it is the highest frequencies where the greatest reduction in the noise level is seen. This is because much of the noise component at low frequencies is made up of the granulation background signal which does not vary significantly with time. In contrast, at high frequencies the granulation background will have dropped almost to zero meaning there will be a greater impact on the noise due to the weighting. Exactly how much of the low frequency background is instrumental and how much is solar is difficult to ascertain. The instrumental noise is clearly not independent of frequency. If one removes the high frequency background from the low frequency background there are still significant variations in the noise amplitude over time. This suggests a higher level of instrumental noise at low frequencies than at high frequencies. The spectral windows, shown in the bottom panels of Fig.~\ref{PowSpecGran}, look similar to those for the simulated data, which is to be expected given the artificial noise was designed to mimic the real noise. While Fig.~\ref{PowSpecGran} shows a reduction in the background for the weighted case, it is not clear how the S/N of the low frequency modes are affected. This is shown more clearly in Fig.~\ref{PowSpecModesGran} where close-up views of the spectra over the region 800 - 1800 $\mu$Hz are given. It can be seen from this plot that the modes in the weighted spectrum have a higher S/N and that there are more visible modes at low frequencies.

\begin{figure}
\centerline{\includegraphics[width=3.0in]{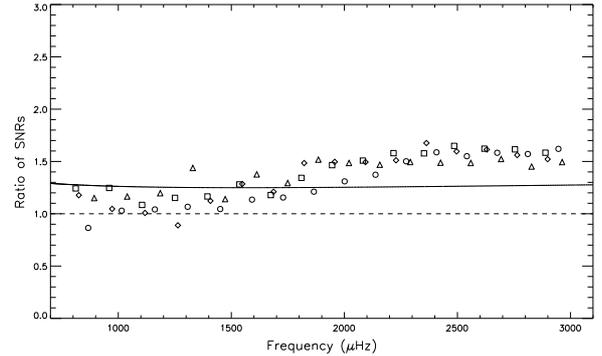}}\caption{Ratio of the S/N for the highest peak in the vicinity of predicted modes frequencies for the real weighted and unweighted data. The symbols represent different $\ell$, with $\ell$=0 given by diamonds, $\ell$=1 by triangles, $\ell$=2 by squares and $\ell$=3 by circles. The solid line shows the `predicted' improvement in S/N (see text) and a dotted line is given at unity to aid the eye.} \label{S/NRGran}
\end{figure}

Fig.~\ref{S/NRGran} shows the ratio of the S/N's for the real BiSON data. They have been calculated using the same smoothing process as described in section~\ref{sec_analysis}. However, since we do not know the true frequencies of the modes in the real data, we compare the S/N's of the maximum peak height in the vicinity of where the modes are predicted to lie. These predictions are based on the estimates returned from model `S' \citep{C-Dalsgaard1996}. Again we see an increase in the S/N when using the weighted data, however this improvement is less than was seen in the simulated case. We also note that the prediction is less than for the simulated case because although the noise characteristics are similar in the two cases they are not identical. With the real data we have plotted down to lower frequencies than for the simulated data (for which we knew there were no modes below 1100 $\mu$Hz). And it seems that even below 1000 $\mu$Hz there is evidence for improvements in S/N for the weighted spectrum. Of course this may just be due to randomly higher noise spikes and so we again need to perform statistical tests to see how likely the peaks are to be due solely to noise.

\begin{table}
\centering \caption{Table giving the frequencies, $\nu$, for the real BiSON data, of the highest peaks after convolution (see section~\ref{sec_analysis}) in regions surrounding the areas where modes are expected to occur for both the unweighted, $u$, and weighted, $w$, data. Also given are the probabilities, $P$, of such peaks occurring, by chance, over a number of bins covering a range of 100 $\mu$Hz. Peaks which are high enough to have only a low probability of occurring by chance ($P<$ 0.1) are printed in bold. Modes which are identified by this means in the weighted spectrum only are denoted by an asterisk. The frequency given for the $\ell$=2, $n$=5 mode denoted by the $\dag$ symbol is likely to be that of the $m$=-2 component only}
\begin{tabular}{ccccccc}
\hline $\ell$ & n & $\nu_u$ ($\mu$Hz) & $P_u$ & $\nu_u$ ($\mu$Hz) & $P_w$ \\
\hline
 0 &  6 &          972.61  &         0.69  &          972.61  &         0.46  &   \\
 0 &  7 &         1119.09  &         1.00  &         1118.20  &         1.00  &   \\
 0 &  8 & \textbf{1263.22} & \textbf{0.00} & \textbf{1263.19} & \textbf{0.01} &   \\
 0 &  9 & \textbf{1407.49} & \textbf{0.00} & \textbf{1407.49} & \textbf{0.00} &   \\
 0 & 10 & \textbf{1548.34} & \textbf{0.00} & \textbf{1548.34} & \textbf{0.00} &   \\
 1 &  6 &         1040.93  &         1.00  &         1038.77  &         0.52  &   \\
 1 &  7 & \textbf{1185.60} & \textbf{0.06} & \textbf{1185.60} & \textbf{0.00} &   \\
 1 &  8 & \textbf{1329.63} & \textbf{0.00} & \textbf{1329.62} & \textbf{0.00} &   \\
 1 &  9 & \textbf{1472.83} & \textbf{0.00} & \textbf{1472.82} & \textbf{0.00} &   \\
 1 & 10 & \textbf{1612.70} & \textbf{0.00} & \textbf{1612.70} & \textbf{0.00} &   \\
 2 &  5 &          961.25  &         0.98  & \textbf{ 958.77} & \textbf{0.03} & *$\dag$ \\
 2 &  6 &         1106.21  &         1.00  &         1104.54  &         0.86  &   \\
 2 &  7 &         1249.50  &         0.98  & \textbf{1251.39} & \textbf{0.02} & * \\
 2 &  8 & \textbf{1394.69} & \textbf{0.00} & \textbf{1394.70} & \textbf{0.00} &   \\
 2 &  9 & \textbf{1535.86} & \textbf{0.00} & \textbf{1535.85} & \textbf{0.00} &   \\
 3 &  5 &         1016.16  &         0.95  &         1013.96  &         0.79  &   \\
 3 &  6 &         1161.26  &         1.00  &         1162.71  &         0.81  &   \\
 3 &  7 &         1308.16  &         0.79  & \textbf{1306.66} & \textbf{0.08} & * \\
 3 &  8 & \textbf{1450.96} & \textbf{0.00} & \textbf{1450.99} & \textbf{0.00} &   \\
 3 &  9 & \textbf{1591.48} & \textbf{0.00} & \textbf{1591.60} & \textbf{0.00} &   \\
\hline \label{TableReal}
\end{tabular}
\end{table}

Table~\ref{TableReal} gives the frequencies of the highest peaks, and as can be seen these values are not necessarily the same in the unweighted as they are in the weighted cases, although where $P<$ 0.1 for both, the values are generally very close. There are three distinct cases where there is a peak within the vicinity of a predicted mode that has a $P$ below the chosen threshold in the weighted spectrum but not in the unweighted one (denoted by an asterisk in Table~\ref{TableReal}). These are the $\ell$=2, $n$=5 and $n$=7 modes and the $\ell$=3, $n=$7 mode. In the case of the $\ell$=2, $n$=5 mode this has never before been identified in the power spectra of any sun-as-a-star data, although given its location, it does seem likely that we have only detected the $m=-2$ component.

\section{Summary}\label{sec_summary}

We have taken a long 3650-day time series of BiSON data and applied a weighting function in order to improve the signal-to-noise ratio (S/N) in p-mode power spectrum. The weighting function was based on the noise levels at low frequencies seen in the power spectra of daily time series from each of the six BiSON stations. A sophisticated simulation was used to track how the noise varies throughout the day, but it was found that including this daily effect in the weighting actually gave little further improvement in the S/N. The sophisticated simulator was also used to create a matching set of artificial data in order to help better understand the improvements that the weighting can bring.

In order to analyse these very long data sets it was necessary to develop a new method for generating the power spectrum of the weighted time series. A traditional sine-wave-fitting method would have taken many weeks to complete for a 3650-day time series, even when using a high performance computing cluster. Instead we developed a method where certain summations used in the sine-wave fitting process are replaced with far more computationally efficient FFT's. This reduced the time to compute the weighted FFT of the 3650-day time series to less than an hour on a single desktop machine.

It was shown that the weighted power spectrum had a significantly lower background level than the unweighted case. However, we also showed that the spectral window of the weighted data was significantly poorer. This means that more power is aliased out of the main central peak into a pseudo-random background structure and prominent sidebands. This is because the application of the weighting function essentially acts as a `dirty' window function. The combined result of these two effects was that we expected an improvement in the S/N of the modes of around 30\% for the real data. By measuring the the actual S/N of the peaks in both the unweighted and weighted spectrum this was shown to be an underestimate for high frequency modes but an overestimate at low frequencies. The actual improvements in the S/N for the low frequency modes seems to be about 15\%-20\%.

Despite this fairly modest improvement there were three more additional peaks that passed the chosen threshold for a statistical test in the weighted data compared with the unweighted data. These were the $\ell$=2, $n$=5 and $n$=7 modes and the $\ell$=3, $n=$7 mode. If confirmed then in the case of the $\ell$=2, $n$=5 and $\ell$=3, $n$=7 modes these would both be new identifications for BiSON data. While the detection of the $\ell$=3, $n$=7 mode was perhaps not unexpected given its expected amplitude, a possible detection of the $\ell$=2, $n$=5 mode is more surprising. The frequency of the peak is about 1$\mu$Hz below what is predicted from the solar models. However this does not rule it out as a potential mode as we could be seeing only the $m=-2$ component. It should be noted that the $\ell$=0, $n$=6 mode which is not seen in either the unweighted or weighted spectra has actually been detected in other earlier BiSON data sets. It seems that this particular mode was excited to greater amplitudes in the past than it is currently and hence tends to appear only in earlier stretches of data.

It is clear that by weighting the BiSON time series we gain a significant improvement in the S/N of the modes and it seems likely that it will be possible to identify more low frequency modes using this technique. We hope to do further studies using this weighting method with different time spans and longer BiSON time series in future. There is also scope for performing a search of contemporaneous BiSON and GOLF data in a similar fashion to that explained in \cite{Broomhall2007}, but using weighted data instead of unweighted.

\renewcommand{\theequation}{A-\arabic{equation}}
\setcounter{equation}{0}
\section*{Appendices}
\subsection{Appendix 1 The SWF formulae}
In this Appendix we present the derivation of Equations 4 and 5 in Section 3. If we fit:
\begin{equation}
A_i \sin(2\pi f_i t_k) + B_i \cos(2\pi f_i t_k),
\end{equation}
to the data, $y_k$, at each frequency, $f_i$, where $t_k$ is the time of the $k$th observation, then the method of least squares \citep[see, for example, page 144 of][]{Bevington1992} allows us to find the values of $A_i$ and $B_i$ by minimising:
\begin{equation}
\sum = \sum (w_k [y_k - A_i sin(2\pi f_i t_k) - B_i cos(2\pi f_i t_k)]^2)
\end{equation}
with respect to $A_i$ and $B_i$. Here $w_k$ represents the statistical weight of the $k$th observation.

It is useful to simplify the notation by setting $s_{ik}=\sin(2\pi f_i t_k)$ and $c_{ik}=\cos(2\pi f_i t_k)$. Taking the partial derivatives of the right hand side of Equation A-2 with respect to $A_i$ and $B_i$ and setting the results equal to zero to identify the minima, we obtain:
\begin{equation}
\frac {\partial \sum}{\partial A_i}=\sum (w_k [- y_k s_{ik}+ A_i s_{ik}^2 + B_i s_{ik} c_{ik}]) = 0,
\end{equation}
and
\begin{equation}
\frac {\partial \sum}{\partial B_i}=\sum (w_k [- y_k c_{ik}+ B_i c_{ik}^2 + A_i s_{ik} c_{ik}]) = 0.
\end{equation}
These two equations may be solved simultaneously to obtain Equations 4 and 5 in Section 3.
\subsection{Appendix 2 Derivation of Equations 6 and 7}

We aim to evaluate the power spectrum of the time series using a sine-wave fit, but with any efficiencies we can use in order to make the calculation less computationally intensive. All the way through this analysis we assume that the time-series has been sampled at a regular cadence (so that its Fast Fourier Transform (FFT) can be calculated) and that the fitted frequencies are those that would have been returned by a FFT of the time-series. So for each frequency, $f_i$, one fits:
\begin{equation}
A_i \sin(2\pi f_i t_k) + B_i \cos(2\pi f_i t_k),
\end{equation}
to the data, $y_k$, where $t_k$ is the time of the $k$th observation.

$A_i$ and $B_i$ can be evaluated by least squares fitting. The appropriate expressions are \citep[Appendix 1;][]{Kjeldsen1992,Frandsen1995}:
\begin{equation}
A_i = \frac {\sum w_k y_k c_{ik} \sum w_k s_{ik} c_{ik} - \sum w_k y_k s_{ik} \sum w_k c_{ik}^2} {(\sum w_k s_{ik} c_{ik})^2 - \sum w_k c_{ik}^2 \sum w_k s_{ik}^2},
\end{equation}
and
\begin{equation}
B_i = \frac {\sum w_k y_k s_{ik} \sum w_k s_{ik} c_{ik} - \sum w_k y_k c_{ik} \sum w_k s_{ik}^2} {(\sum w_k s_{ik} c_{ik})^2 - \sum w_k c_{ik}^2 \sum w_k s_{ik}^2},
\end{equation}
where $s_{ik}=\sin(2\pi f_i t_k)$, $c_{ik}=\cos(2\pi f_i t_k)$ and $w_k$ is the statistical weight of the $k$th measurement.

The power spectrum can then be obtained by evaluating $A_i^2 + B_i^2$  as a function of $i$. This takes significantly longer to compute than, say, a power spectrum that uses an FFT approach, because each unique summation in the expressions for $A_i$ and $B_i$ must be evaluated for each frequency, i.e, for each value of $i$.

It is, however, possible to make use of Fourier Transform properties so that the summations required can be evaluated using FFT methods, thus speeding up the process of sine-wave fitting. To see how this works, we start from the definition of the Fourier Transform and then identify its real and imaginary parts with integrals over sines and cosines.

The Fourier transform, $G(f)$, of a function of time, $g(t)$, is defined by:
\begin{eqnarray}
G(f) & = & \int_{-\infty}^\infty g(t)e^{-2\pi i f t}dt \\
     & = & \int_{-\infty}^\infty g(t)\cos(2\pi i f t)dt \nonumber  - j\int_{-\infty}^\infty g(t)\sin(2\pi i f t)dt.
\end{eqnarray}
If g(t) represents a set of measurements (i.e., $g(t)$ is always real), then the two integrals above are always real and we can write:
\begin{equation}
G(f) = \textrm{Re}[G(f)] + j \textrm{ Im}[G(f)],
\end{equation}
where
\begin{equation}
\textrm{Re}[G(f)] = \int_{-\infty}^\infty g(t)\cos(2\pi i f t)dt,
\end{equation}
and
\begin{equation}
\textrm{Im}[G(f)] = -\int_{-\infty}^\infty g(t)\sin(2\pi i f t)dt.
\end{equation}
$G(f)$ can be evaluated using FFT methods and the integrals above can be replaced by summations if $g(t)$ is a discrete function of time. To link this explicitly to our situation, we can define the $k$th element of $g$ by:
\begin{equation}
g(t_k) = w_k y_k,
\end{equation}
and therefore two of the summations we need to evaluate (see Equation A-6 and A-7) are given by:
\begin{equation}
\sum w_k y_k c_{ik} = N \textrm{Re}[\textrm{FFT}(w_k y_k)],
\end{equation}
and
\begin{equation}
\sum w_k y_k s_{ik} = - N \textrm{Im}[\textrm{FFT}(w_k y_k)],
\end{equation}

The numerical factor of $N$ arises in Equations A-13 and A-14 if we adopt the definition of the FFT used, for example, by the IDL language, i.e. if the FFT of $x(t)$ is defined by:
\begin{eqnarray}
\textrm{FFT}(x(t))& = & \frac{1}{N}\sum (x(t)e^{-2\pi j f t})\\
                  & = & \textrm{Re}[\textrm{FFT}(x(t))] + j \textrm{Im}[\textrm{FFT}(x(t))]\nonumber
\end{eqnarray}
The remaining three summations in Equations A-6 and A-7 can easily be approximated. Firstly, the summations over sc, s$^2$ and c$^2$ can be expressed in terms of sinusoidal functions of the double angle, i.e.,
\begin{equation}
\sum w_k s_{ik} c_{ik} = \sum \frac{w_k}{2} \sin(2\times2\pi f_i t_k),
\end{equation}
\begin{equation}
\sum w_k c_{ik}^2 = \sum \frac{w_k}{2} (1+\cos(2\times2\pi f_i t_k)),
\end{equation}
\begin{equation}
\sum w_k s_{ik}^2 = \sum \frac{w_k}{2} (1-\cos(2\times2\pi f_i t_k)),
\end{equation}
Secondly, we note that the the summation in Equation A-16 and the second terms in the summations of Equations A-17 and A-18 essentially ``pick out" the harmonic content of $w_k$. They will be very nearly zero when the frequency under consideration is well away from any frequency content of $w_k$. This is true in the BiSON case, where the changes in $w_k$ are very much slower than changes in $y_k$ due to solar p modes. This means that for all frequencies of interest the first terms in each of Equations A-17 and A-18 will be dominant and constant. So to a very good approximation, Equations A-6 and A-7 can be re-written:-
\begin{equation}
A_i = \frac{-N \textrm{Im}[\textrm{FFT}(w_k y_k)]}{\sum(w_k/2)}
\end{equation}
and
\begin{equation}
B_i = \frac{N \textrm{Re}[\textrm{FFT}(w_k y_k)]}{\sum(w_k/2)}.
\end{equation}

These are Equations 6 and 7 in the main text. In this approximation we see that the usual method of calculating the power spectrum after performing a SWF, i.e. by evaluating $A_i^2 + B_i^2$, yields the same result as taking the modulus-squared of the FFT of $w_k y_k$ (to within a numerical multiplier of $4 N^2/(\Sigma w_k)^2$). The multiplier equals 4 in the unweighted case (i.e. where all the $w_k$ equal unity). The discussions of Section 5 rely on the fact that in the weighted case we are essentially dealing with a time series given by $w_k y_k$.

\subsection{Appendix 3 Limiting value of Equation 20}
We begin by noting that the product of the summations has $N^2$ terms, $N$ of which are just given by $w_k/w_k$ (each of which is equal to 1 of course) and $N^2-N$ of which are $w_j/w_k$. The latter terms can be ``paired up", so that there are $(N^2-N)/2$ terms of the kind $w_j/w_k + w_k/w_j$. If each of these terms is greater than 2, then the product of $\sum{\frac{1}{w_k}}\sum{w_j}$ will be greater than $N^2$. To show that this is indeed the case, we begin by defining;
\begin{equation}
x_{jk}=w_j/w_k,\label{xDef}
\end{equation}
so that each paired term, $t_{jk}$, is given by:
\begin{equation}
t_{jk}=x_{jk}+\frac{1}{x_{jk}}.\label{tDef}
\end{equation}

By differentiating $t$ with respect to $x$ it can easily be shown that $t_{jk}$ has a minimum value of 2 when $x_{jk} = 1$. Hence:
\begin{equation}
\sum{\frac{1}{w_k}}\sum{w_j}= N^2,
\end{equation}
if all the $x_{jk}$ are equal to 1, i.e. if all the data points have equal weight, and is otherwise greater than $N^2$, in which case the LHS of Eqn.~\ref{S/Nratio} is greater than 1.

\section*{Acknowledgments}

We would like to thank all those who are, or have been, associated
with BiSON. BiSON is funded by the Science and Technology Facilities
Council (STFC).
We should also like to thank the referees for valuable comments that have allowed us to improve the paper.

\bibliography{fletcher}

\def\noopsort#1{}
\begin{thebibliography}{}

\bibitem[\protect\citeauthoryear{{Appourchaux}}{{Appourchaux}}{2004}]{Appourch%
aux2004}
{Appourchaux} T.,  2004, A\&A, 428, 1039

\bibitem[\protect\citeauthoryear{{Bevington} \& {Robinson}}{{Bevington}~\&~Robinson}{1992}]{Bevington1992}
{Bevington} P.~R., {Robinson} D.~K.,  1992, {McGraw-Hill}, {Data Reduction And Error Analysis For The Physical Sciences}

\bibitem[\protect\citeauthoryear{{Brigham}}{{Brigham}}{1988}]{Brigham1988}
{Brigham} E.~O.,  1988, {Prentice Hall}, {The Fast Fourier Transform And Its Applications}

\bibitem[\protect\citeauthoryear{{Brookes}, {Isaak} \& {van der
  Raay}}{{Brookes} et~al.}{1978}]{Brookes1978}
{Brookes} J.~R.,  {Isaak} G.~R.,    {van der Raay} H.~B.,  1978, MNRAS, 185, 1

\bibitem[\protect\citeauthoryear{{Broomhall}, {Chaplin}, {Elsworth} \&
  {Appourchaux}}{{Broomhall} et~al.}{2007}]{Broomhall2007}
{Broomhall} A.~M.,  {Chaplin} W.~J.,  {Elsworth} Y.,    {Appourchaux} T.,
  2007, MNRAS, 379, 2

\bibitem[\protect\citeauthoryear{{Broomhall}, {Chaplin}, {Elsworth} \&
  {New}}{{Broomhall} et~al.}{2009}]{Broomhall2009a}
{Broomhall} A.~M.,  {Chaplin} W.~J.,  {Elsworth} Y.,    {New} R.,  2009, MNRAS,
  397, 793

\bibitem[\protect\citeauthoryear{{Chaplin}, {Elsworth}, {Isaak}, {Marchenkov},
  {Miller}, {New}, {Pinter} \& {Appourchaux}}{{Chaplin}
  et~al.}{2002}]{Chaplin2002}
{Chaplin} W.~J.,  {Elsworth} Y.,  {Isaak} G.~R.,  {Marchenkov} K.~I.,  {Miller}
  B.~A.,  {New} R.,  {Pinter} B.,    {Appourchaux} T.,  2002, MNRAS, 336, 979

\bibitem[\protect\citeauthoryear{{Chaplin}, {Elsworth}, {Isaak}, {Miller},
  {New} \& {Pint{\'e}r}}{{Chaplin} et~al.}{2005}]{Chaplin2005a}
{Chaplin} W.~J.,  {Elsworth} Y.,  {Isaak} G.~R.,  {Miller} B.~A.,  {New} R.,
  {Pint{\'e}r} B.,  2005, MNRAS, 359, 607

\bibitem[\protect\citeauthoryear{{Christensen-Dalsgaard}, {et al.} \&
  {}}{{Christensen-Dalsgaard} et~al.}{1996}]{C-Dalsgaard1996}
{Christensen-Dalsgaard} J.,  {et al.}   {} 1996, Sci, 272, 1286

\bibitem[\protect\citeauthoryear{{Elsworth}, {Howe}, {Isaak}, {McLeod},
  {Miller}, {New} \& {Wheeler}}{{Elsworth} et~al.}{1995}]{Elsworth1995}
{Elsworth} Y.,  {Howe} R.,  {Isaak} G.~R.,  {McLeod} C.~P.,  {Miller} B.~A.,
  {New} R.,    {Wheeler} S.~J.,  1995, A\&AS, 113, 379

\bibitem[\protect\citeauthoryear{{Fletcher} S.~T.~{New}, {Broomhall}, {Chaplin} \&
  {Elsworth}}{{Fletcher} et~al.}{2009}]{Fletcher2009c}
{Fletcher} S.~T.~, {New}
  R., {Broomhall} A.-M.,  {Chaplin} W.~J.,  {Elsworth} Y., 2009, in {Dikpati} M.,  {Gonzalez-Hernandez} I.,  {Artentoft} T.,
  {Hill} F.,  eds, ASP: GONG 2008/SOHO XXII meeting on solar-stellar dynamos as
  revealed by Helio- and Asteroseismology, Vol. 416, 337

\bibitem[\protect\citeauthoryear{{Frandsen}, {Jones}, {Kjeldsen}, {Viskum},
  {Hjorth}, {Andersen} \& {Thomsen}}{{Frandsen} et~al.}{1995}]{Frandsen1995}
{Frandsen} S.,  {Jones} A.,  {Kjeldsen} H.,  {Viskum} M.,  {Hjorth} J.,
  {Andersen} N.~H.,    {Thomsen} B.,  1995, A\&A, 301, 123

\bibitem[\protect\citeauthoryear{{Harvey}}{{Harvey}}{1985}]{Harvey1985}
{Harvey} J.,  1985, Technical report, {High-resolution helioseismology}

\bibitem[\protect\citeauthoryear{{Jim{\'e}nez-Reyes} \&
  {et~al.}}{{Jim{\'e}nez-Reyes} \& {et~al.}}{2008}]{Jimenez2008}
{Jim{\'e}nez-Reyes} S.~J.,  {et~al.} 2008, MNRAS, 389, 1780

\bibitem[\protect\citeauthoryear{{Kjeldsen}}{{Kjeldsen}}{1992}]{Kjeldsen1992}
{Kjeldsen} H., 1992, Ph.D. Thesis, University of Aarhus, Denmark

\bibitem[\protect\citeauthoryear{{New}, {Fletcher}, {Chaplin} \&
  {Elsworth}}{{New} et~al.}{2009}]{New2009}
{New} R.,  {Fletcher} S.~T.,  {Chaplin} W.~J.,    {Elsworth} Y.,  2009, in
  {Dikpati} M.,  {Gonzalez-Hernandez} I.,  {Artentoft} T.,   {Hill} F.,  eds,
  ASP: GONG 2008/SOHO XXI meeting on solar-stellar dynamos as revealed by
  Helio- and Asteroseismology, Vol. 416, 325

\bibitem[\protect\citeauthoryear{{Press}, {Teukolsky}, {Vetterling} \& {Flannery}}{{Press} et~al.}{2007}]{Press2007}
{Press} W.~H., {Teukolsky} S.~A., {Vetterling} W.~T., {Flannery B.~P.},  2007, {Cambridge University Press}, %
 {Numerical Recipes; The Art of Scientific Computing}

\bibitem[\protect\citeauthoryear{{Salabert}, {Leibacher} \&
  {Appourchaux}}{{Salabert} et~al.}{2008}]{Salabert2008}
{Salabert} D.,  {Leibacher} J.~W.,    {Appourchaux} T.,  2008, Journal of
  Physics Conference Series, 118, 012086

\bibitem[\protect\citeauthoryear{{Scargle}}{{Scargle}}{1982}] {Scargle1982}
  {Scargle} J.~D. 1982, ApJ, 263, 835


\end{thebibliography}

\end{document}